\documentclass[pre,superscriptaddress,showpacs,twocolumn]{revtex4}
\usepackage{graphicx,amsfonts}
\usepackage{epsfig,amsmath,eufrak}

\usepackage{natbib,layout}

\begin{document}

\newcommand{\atanh}
{\operatorname{atanh}}
\newcommand{\ArcTan}
{\operatorname{ArcTan}}
\newcommand{\ArcCoth}
{\operatorname{ArcCoth}}
\newcommand{\Erf}
{\operatorname{Erf}}
\newcommand{\Erfi}
{\operatorname{Erfi}}
\newcommand{\Ei}
{\operatorname{Ei}}

\title{Universal aging properties at a disordered critical point} 
\author{Gregory Schehr}
\author{Raja Paul}

\affiliation{Theoretische Physik, Universit\"at des Saarlandes,
66041 Saarbr\"ucken, Germany}

\draft

\date{\today}
\begin{abstract}
We investigate, analytically near the dimension $d_{uc}=4$ and
numerically in $d=3$, the non equilibrium relaxational dynamics of the
randomly diluted Ising model at criticality. 
Using the Exact Renormalization Group Method to
one loop, we compute the two times $t,t_w$ correlation function and Fluctuation
Dissipation Ratio (FDR) for any Fourier mode of the order parameter,
of finite wave vector $q$.  
In the large time separation limit, the FDR is found to reach
a non trivial value $X^{\infty}$ independently of (small) $q$
and coincide with the FDR associated to the 
the {\it total} magnetization obtained previously. Explicit
calculations in real 
space show that the FDR associated to the {\it local} magnetization 
converges, in the asymptotic limit, to this same value
$X^{\infty}$. Through a Monte Carlo simulation, we compute
the autocorrelation function in three dimensions, for different values
of the dilution fraction $p$ at $T_c(p)$. Taking
properly into account the corrections to scaling, we find, according
to the Renormalization Group predictions, that 
the autocorrelation exponent $\lambda_c$ is independent on $p$. 
The analysis is complemented by a study of the non equilibrium 
critical dynamics following a quench from a completely ordered state.    
\end{abstract}
\maketitle

The study of relaxational dynamics following a quench at a pure
critical point has attracted much attention these last few years
\cite{bray_domaingrowth_review, cugliandolo_fdr_pure,
  godreche_crit_ferro_review, calabrese_review_fdr}.   
Although simpler to study than glasses, 
critical dynamics display interesting non equilibrium
features such as aging, commonly observed in more complex disordered or  
glassy phases \cite{cugliandolo_leshouches}. In this context, the
computation of two times $t,t_w$ response and correlation functions
with associated universal exponents has been the subject of numerous
analytical as well as numerical studies \cite{calabrese_review_fdr}.  

In addition, it has been proposed \cite{godreche_fdr_crit_ferro} that
a non trivial Fluctuation 
Dissipation Ratio (FDR) $X$, originally introduced in the 
Mean Field approach to glassy systems, which  
generalizes the Fluctuation Dissipation Theorem (FDT) to non
equilibrium situations, is a new {\it universal} quantity
associated to these critical points. As such, it has been computed using the
powerful tools of RG, {\it e.g.} for pure $O(N)$ model at
criticality in the vicinity of the upper critical 
dimension $d_{uc} = 4$ and for various dynamics
\cite{calabrese_review_fdr,
calabrese_on_oneloop,calabrese_on_conserv}.    

An important question related to the physical interpretation of $X$ in
terms of
an effective temperature \cite{leticia_teff} $T_{\text{eff}} = T/X$ is 
its dependence on the observables \cite{calabrese_eff_temp,mayer_fdr}. 
In this respect, a heuristic argument \cite{calabrese_on_oneloop}
suggests that, for a wide class of critical systems, the {\it local}
FDR associated 
to correlation and corresponding response of the local magnetization
should be identical, in the large time separation limit, to
the FDR for the {\it total} magnetization, {\it i.e.} for the Fourier mode
$q=0$. This argument relies strongly on the hypothesis that the
time decay of the response function of the Fourier mode $q$ is
characterized by a single time scale $\tau_q \sim q^{-z}$, with $z$ the
dynamical exponent. 

Characterizing the effects of quenched disorder on
critical dynamics is a complicated task and indeed  
the question of how quenched randomness modifies these 
properties has been much less studied. In particular, in this context
of critical disordered systems, the question of
universality, {\it i.e.} the dependence of the critical exponents on
the strength of the disorder, is a controversial issue \cite{heuer_tc}.   
In this paper, we address these questions   
on the randomly diluted Ising model: 
\begin{eqnarray}\label{def_diluted}
H = \sum_{\langle i j \rangle} \rho_i \rho_j s_i s_j
\label{eq_Hamil}
\end{eqnarray}  
where $s_i$ are Ising spins on a $d$-dimensional hypercubic lattice
and $\rho_i = 1$ with probability $p$ and
$0$ with probability $1-p$. For the experimentally relevant case of
dimension $d=3$ \cite{belanger_exp}, for which the specific heat
exponent of the pure 
model is positive, the disorder is expected, according to Harris
criterion\cite{harris_criterion}, to modify the universality class of the
transition. For $1-p \ll 1$,  the
large scale properties of (\ref{def_diluted}) at criticality
are then described by the
following $\text{O}(1)$ model with a random mass term, the so-called Random
Ising Model (RIM): 
\begin{equation}\label{H_rim}
H^{\psi}[\varphi] = \int d^d x \left[ \frac{1}{2}(\nabla \varphi)^2 +
 \frac{1}{2} [r_0 + \psi(x)] \varphi^2 + \frac{g_0}{4!}
 \varphi^4 \right]  
\end{equation} 
where $\varphi \equiv \varphi(x)$ and  
$\psi(x)$ is a gaussian random variable $\overline{\psi(x)
  \psi(x')} = \Delta \delta^d(x-x')$ and $r_0$, the bare mass, is
  adjusted so that the renormalized one is zero. The static critical
  properties of this model have been intensively 
studied \cite{folk_rim_review} both
analytically, mainly using RG, within various schemes, and
numerically \cite{ballesteros_tc}. The (perturbative)
RG calculations below the upper critical dimension in $d=4-\epsilon$,
which we will focus on here, confirm the qualitative Harris
criterion and predict that the critical properties of these
models (\ref{H_rim}) for different
values of $p$ close to one, 
are described by a new disordered fixed point, which is 
independent of $p$. Therefore, an important statement of this RG
analysis is that the critical exponents, which can be
computed in an expansion of $\sqrt{\epsilon}$, {\it e.g.} $\eta =
{\cal O}(\epsilon)$, are universal, {\it i.e.} 
{independent} of $p$. 
This was recently confirmed by Monte Carlo simulation in
$d=3$\cite{ballesteros_tc}, over a wide range of concentration $p$
above the percolation threshold $p_c=0.31$. Although quantitative
discrepancies were 
found with perturbative RG 
calculations \cite{foot_borel}, universality was demonstrated by
taking carefully into 
account the (strong) corrections to scaling \cite{ballesteros_tc}.    
In the equilibrium dynamics, at variance
with the pure case, the perturbative expansion of the
dynamical exponent $z$ differs from its high temperature value of $2$
already at one loop \cite{grinstein_randommass_z_oneloop} $z - 2 =
\sqrt{\frac{6\epsilon}{53}} + {\cal 
  O}(\epsilon)$ independently of $p$, corrections 
to two loops have been computed in Ref.
\cite{janssen_randommass_z_twoloops} and up to three loops in Ref. 
\cite{prudni_three_loops}. After a long debate, 
a recent numerical simulation\cite{parisi_simu_rim} where
corrections to (dynamical) scaling were taken into account, has
confirmed the universality of $z$ in $d=3$, leading to $z = 2.62(7)$
independently on the spin concentration $p$ above $p_c$. 

By contrast, much less is known about the non equilibrium dynamics of
this disordered system at criticality. The critical initial slip
exponent $\theta'$ vanishes to one loop \cite{kissner_random_mass}, 
and correction to two-loops have been computed
\cite{oerding_randommass_theta_twoloops}. This exponent has been
recently 
computed up to two loops for the case of extended defects in
\cite{andrei_ext}. The two times response
in Fourier space, ${\cal R}^q_{tt_w}$ \cite{kissner_random_mass} and the
correlation  
${\cal C}^{q=0}_{tt_w}$ \cite{calabrese_fdr_randommass}, including the
associated scaling functions, are known up to one
loop. But although the dynamical RG predicts a 
universal value of the autocorrelation exponent $\lambda_c = d
-z\theta'$, 
for $p$ close to $1$, this 
statement remains an open question
for a wider range of values of $p$. Furthermore, a non trivial  
FDR \cite{calabrese_fdr_randommass}, only for the total magnetization,
was recently obtained to one loop, and 
it was argued,  
using the same aforementioned heuristic argument
\cite{calabrese_on_oneloop}, to 
coincide with the {\it local} FDR.   
However, it was already noticed in Ref.\cite{kissner_random_mass}
that, due to the disorder,  
${\cal R}^q_{tt_w}$ decays as a {\it power law} for $q^zt \gg
1$. Therefore, 
the argument of Ref. \cite{calabrese_on_oneloop} is challenged for
this disordered critical point, and, already at one loop order,  
the computation of the FDR needs a
closer inspection, including an extension of the analysis of Ref. 
\cite{calabrese_fdr_randommass} beyond the ``diffusive'' $q=0$ mode.

In this paper, using RG to one loop, we obtain, for any finite
Fourier mode $q$, the correlation function ${\cal C}^q_{tt_w}$
and the FDR  ${
  X}^q_{tt_w}$, which are both characterized by scaling functions of the
variables $q^z(t-t_w)$ and $t/t_w$. In the asymptotic large time
separation regime $t \gg t_w$, the FDR reaches a non trivial value
$X^{\infty}$, {\it independently} of (small) $q$. In addition, we 
explicitly compute the {\it local} FDR, which is a function of $t/t_w$ and
reaches the same non-trivial limit $X^{\infty}$ when $t \gg  t_w$,
which thus establishes on firmer ground
the heuristic argument of Ref.\cite{calabrese_on_oneloop}
for the present disordered case. Besides, we perform a Monte Carlo
simulation of the non-equilibrium relaxation of (\ref{def_diluted}) 
following a quench from high temperature with initial magnetization $m_0=0$ 
at $T_c(p)$ and compute the autocorrelation function. In the asymptotic
regime, it takes a scaling form compatible with the RG calculations. By
taking into account corrections to scaling, we show
that the exponent $\lambda_c$ is independent of $p$. Finally, we compute
numerically the autocorrelation function for the critical dynamics
following a quench for a completely ordered initial condition with 
$m_0 = 1$. We observe that the system is also aging and 
show that the decaying exponent $\lambda_c$ is 
strongly affected by this initial condition. 

We study the relaxational dynamics of the Randomly diluted Ising Model
in dimension $d=4-\epsilon$ described by a Langevin equation:
\begin{eqnarray}\label{def_Langevin}
\eta \frac{\partial}{\partial t} \varphi(x,t) = - \frac{\delta
 H^{\psi}[\varphi]}{\delta \varphi(x,t)} + \zeta(x,t)
\end{eqnarray}
where where $\langle \zeta(x,t) \rangle = 0$ and
$\langle \zeta(x,t) \zeta(x',t') \rangle = 2\eta T \delta(x-x')
\delta(t-t') $ is the thermal noise and $\eta$ the friction
coefficient. At initial time $t_i = 0$, the system is in a random
initial configuration with {\it zero} magnetization $m_0=0$
distributed according to 
a Gaussian with short range correlations 
\begin{eqnarray}\label{ini_cond}
[\varphi(x,t=0)\varphi(x',t=0)]_i = \tau_0^{-1} \delta^{d}(x-x')
\end{eqnarray}
Notice that it has been shown that $\tau_0^{-1}$ is here 
irrelevant (in the RG sense) in the large time regime studied here
\cite{janssen_gen}. We will focus on the correlation ${\cal
C}^q_{tt_w}$ in Fourier space and the autocorrelation $C_{tt_w}$  
\begin{eqnarray}\label{def_C}
{\cal{C}}^q_{tt_w} =\overline{\langle {\varphi}(q,t)
      {\varphi}(-q,t_w)  \rangle} \quad, \quad  C_{tt_w} =
      \overline{\langle {\varphi}(x,t) 
      {\varphi}(x,t_w)  \rangle}
\end{eqnarray}
and the response ${\cal R}^q_{tt_w}$ to a small external
field ${f}(-q,t_w)$ as well as on the local response function
${R}^{}_{tt_w}$ respectively defined, for $t>t_w$   
\begin{eqnarray}\label{def_R}
{\cal R}^q_{tt_w} = \overline{\frac{\delta \langle
  {\varphi}(q,t) \rangle}{\delta 
  {f}(-q,t_w)} } \quad, \quad {R}^{}_{tt_w} 
 = \overline{\frac{\delta \langle {\varphi}(x,t) \rangle}{\delta 
  {f}(x,t_w)} }
\end{eqnarray}
where $\overline{..}$ and $\langle .. \rangle $
denote respectively averages w.r.t. disorder and 
thermal fluctuations. We focus also on the 
FDR ${X}^q_{tt_w}$  associated to 
the observable $\varphi$ \cite{cugliandolo_leshouches}:
\begin{equation}
\frac{1}{{X}^q_{tt_w}} = \frac{\partial_{t_w} {\cal
    C}^q_{tt_w}}{T {\cal R}^q_{tt_w}}
 \label{def_FDR}
\end{equation}
defined such that ${{X}^q_{tt_w}} = 1$ at equilibrium. Notice also
that for this choice of initial conditions (\ref{ini_cond}), connected and
non connected correlations do coincide for large system size.  

A convenient way to study the Langevin dynamics defined by
Eq. (\ref{def_Langevin}) is to use the
Martin-Siggia-Rose generating functional. Using the Ito
prescription, it can be readily averaged over the disorder.
The correlations (\ref{def_C}) and response
(\ref{def_R}) are then obtained from a dynamical (disorder averaged)
generating functional or, equivalently, as
functional derivatives of the corresponding dynamical {\it effective}
action $\Gamma$. This functional can be perturbatively computed
\cite{schehr_co_pre} 
using the Exact RG equation associated to the
multi-local operators expansion introduced in
\cite{chauve_erg,scheidl_multilocal}. 
It allows to handle arbitrary cut off functions $c(q^2/2 \Lambda_0^2)$
and check 
universality, independence w.r.t. $c(x)$ and the ultraviolet scale
$\Lambda_0$.  
It describes the evolution of $\Gamma$
when an additional infrared cut off $\Lambda_l$ is lowered from 
$\Lambda_0$ to its 
final value $\Lambda_l \to 0$ where a fixed point of order
${\cal O}(\sqrt{\epsilon})$ is reached. In this
limit, one obtains ${\cal R}^q_{tt_w}$ and ${\cal C}^q_{tt_w}$
(for $t>t_w$) from
\begin{equation}
\partial_{t} {{\cal R}}_{tt_w}^{q} + (q^2 + \mu(t))
{{\cal R}}_{tt_w}^{q} +
\int_{t_i}^t d {t_1} \Sigma_{tt_1}
{{\cal R}}_{t_1t_w}^{q} = 0   \label{Eq_R}
\end{equation}
\begin{equation}
{\cal C}^{ {q}}_{ {t} {t_w}} =
2T\int_{t_i}^{{t_w}} dt_1 {{\cal R}}^{ {q}}_{ {t}t_1}{{\cal
R}}^{ {q}}_{ {t_w}t_1} 
+ \int_{t_i}^{ {t}} dt_1 \int_{t_i}^{ {t_w}} dt_2{{\cal
R}}^{ {q}}_{ {t} t_1} D_{t_1t_2}
{{\cal R}}^{ {q}}_{ {t_w} t_2}   
\label{Eq_C}
\end{equation}
with $\mu(t) = - \int_{t_i}^t d {t_1} \Sigma_{tt_1}$ and 
where the self energy $\Sigma_{t_1 t_2}$ and the noise-disorder kernel 
$D_{t_1 t_2}$ are directly obtained
from $\Gamma$ at the fixed point. One finds:
\begin{eqnarray}
&& \Sigma_{t t'} = -\frac{1}{2} \sqrt{\frac{6\epsilon}{53}} \int_a
  (\gamma_a(t-t'))^2 \\ 
&& D_{t t'} = \frac{T_c}{2} \sqrt{\frac{6\epsilon}{53}} \int_a
  \left( \gamma_a(t-t') - \gamma_a(t+t')\right)
\end{eqnarray} 
where $\gamma_a(x) = (x+a/(2\Lambda_0^2))^{-1}$. For concrete calculations, we 
have used the decomposition of the cut-off funtion 
$c(x) = \int da \hat{c}(a) e^{-a x} \equiv \int_a e^{-a x}$.  

The computation of the correlation function ${\cal C}^q_{tt_w}$
requires the knowledge of the response, which we first focus on. 
By solving perturbatively to order ${\cal O}(\sqrt{\epsilon})$ the
differential equation (\ref{Eq_R}), similarly to what is done in
Ref. \cite{schehr_co_pre}, one recovers, in the limit
$q/\Lambda_0 \ll 1$ keeping the scaling variables $v=q^z(t-t_w)$ and
$u=t/t_w$ fixed, the solution obtained in
Ref. \cite{kissner_random_mass}, consistent with the 
scaling form 
\begin{eqnarray}
&&{\cal R}^{ {q}}_{ {t} {t_w}} =
 {q}^{-2+z+\eta}\left( \frac{ {t}}{ {t_w}} \right)^{\theta}
F_R( {q}^z( {t}- {t_w}), {t}/ {t_w})
\label{janssenscalingresp}
\end{eqnarray}
where $\theta =  \frac{1}{2} \sqrt{\frac{6\epsilon}{53}} + {\cal
  O}(\epsilon)$ and the universal \cite{foot_univ} scaling function $F_R(v,u)$
admits also an
expansion in powers of $\sqrt{\epsilon}$
with \cite{kissner_random_mass} $F_R(v,u) \equiv F_R^{\text{eq}}(v) =
  e^{- v} + \frac{1}{2}  
\sqrt{\frac{6\epsilon}{53}} ((v-1) \Ei{(v)} e^{- v} + e^{- v} - 1)$
where $\Ei(v)$ is the exponential integral function. At variance with
  the pure model at one loop \cite{calabrese_on_oneloop}, the large $v$
  behavior of $F_R^{\text{eq}}(v)$ is a power law, $F_R(v)\propto
  v^{-2}$, which already indicates that the heuristic argument of
  Ref.\cite{calabrese_on_oneloop} can not be applied here. Besides, 
when computing the local response
${R}^{}_{tt_w}$, one is left with an integral over momentum
which is logarithmically divergent, indicating that this integral has
to be handled with care to obtain the correct result, as the scaling
form in (\ref{janssenscalingresp}) is valid only for $q/\Lambda_0 \ll
1$. We thus solve perturbatively the equation (\ref{Eq_R}) for any
  fixed $q$ and obtain an expression for ${\cal R}^q_{tt_w}$
  consistent with the scaling form
\begin{eqnarray}\label{resp_full_q}
{\cal R}^q_{tt_w} = {\tilde g}_1(q) \left( \frac{ {t}}{ {t_w}} \right)^{\theta}
F_R( {\tilde g}_2(q) {q}^2( {t}- {t_w}), {t}/ {t_w})
\end{eqnarray}
where ${\tilde g}_1(q), {\tilde g}_2(q)$ are non-universal functions,
{\it i.e.} which depend explicitly on the cut-off function $c(x)$ and
$\Lambda_0$, 
with the universal \cite{foot_univ} small $q$ behavior:
\begin{eqnarray}
{\tilde g}_1(q) \sim q^{z-2+\eta} \quad, \quad {\tilde g}_2(q) \sim
q^{z-2} 
\end{eqnarray}
which thus allows to recover the previous expression in the asymptotic limit
$q/\Lambda_0 \ll 1$ (\ref{janssenscalingresp}). By computing the
Fourier transform of ${\cal R}^q_{tt_w}$ as given in
Eq. (\ref{resp_full_q}), we
explicitly check that the local
response $R_{tt_w}$ is consistent with the scaling form
\begin{eqnarray}\label{scal_auto_resp}
{R}^{}_{tt_w} = \frac{K_d}{2}
\frac{A^0_{\cal R} + A^1_{\cal R}
\ln{(t-t_w)}}{(t-t_w)^{1+(d-2+\eta)/z}}\left(\frac{t}{t_w}\right)^{\theta}  
\end{eqnarray}
with $K_d=S_d/(2\pi)^d$ and
where the non-universality is left in the
amplitudes $A^0_{\cal R}$ and $A^1_{\cal R}$:
\begin{eqnarray}
&&A^0_{R} = 1 - \frac{3}{2}\sqrt{\frac{6\epsilon}{53}} + \rho_{
 R} \quad, \quad A^1_{R} = \frac{1}{2}\sqrt{\frac{6\epsilon}{53}}    \\
&&\rho_{R} = \sqrt{\frac{6\epsilon}{53}}\int_a
  \ln{\left(\frac{2\Lambda_0^2}{a}\right)} \nonumber 
\end{eqnarray}
At the order of our calculations ${\cal O}(\sqrt{\epsilon})$, 
although $z \neq 2$, this scaling form (\ref{scal_auto_resp}) is
compatible with local scale invariance arguments \cite{henkel_lsi}. 
Notice also that, at this order, the scaling form obtained for 
${R}^{}_{tt_w}$ could be written as
\begin{eqnarray}\label{alter_scal_auto_resp}
{R}^{}_{tt_w} =  \frac{K_d}{2}A_{R}
\frac{1}{(t-t_w)^{(1+a)}} \left(\frac{t}{t_w}\right)^{\theta}
\end{eqnarray}
with $a \neq (d-2+\eta)/z$. Although this scaling form
(\ref{alter_scal_auto_resp}) can not be ruled out at 
this stage, which would in principle require a $2$-loop calculations, 
it seems
rather unlikely given the scaling form obtained in Fourier space
(\ref{janssenscalingresp}), where instead a logarithmic correction
as in (\ref{scal_auto_resp}) is suggested by the large argument
behavior of the function $F_R^{\text{eq}}(v)$.   

We now turn to the computation of the correlation function in Fourier space
${\cal C}^q_{tt_w}$, which was only computed for the $q=0$ mode
\cite{calabrese_fdr_randommass}. Solving 
(\ref{Eq_C}), one obtains an explicit expression, which, in the
aforementioned scaling limit, is compatible with the scaling form
\begin{eqnarray}
&&{\cal C}^{{q}}_{{t}{t_w}} = T_c
{q}^{-2+\eta}\left( \frac{{t}}{{t_w}} \right)^{\theta}
F_C({q}^z({t}-{t_w}),{t}/{t_w})
\label{janssenscalingcorr}
\end{eqnarray}
with the full expression 
\begin{eqnarray}\label{perturb_correl}
&&F_C(v,u) = F_C^0(v,u) + \sqrt{\epsilon} F_C^1(v,u) + {\cal O}(\epsilon) \\
&&F_C^0(v,u) = e^{-v} - e^{-v(u+1)/(u-1))} \nonumber \\
&&F_C^1(v,u) = F_C^{1 \text{eq}}(v) - F_C^{1 \text{eq}}{\left(v
    \frac{u+1}{u-1}\right)}  \\
&&+\sqrt{\frac{6}{53}} e^{-v \frac{u+1}{u-1}}\left( 
  \Ei{\left(\frac{2v}{u-1}\right)} -  \ln{\left(\frac{2v}{u-1}\right)}
  - \gamma_E  
\right) \nonumber \\
&&F_C^{1 \text{eq}}(v) = \frac{1}{2}\sqrt{\frac{6}{53}}\left( 
e^{-v} + v e^{-v} \Ei{(v)} \right)
\end{eqnarray} 
where $\gamma_E$ is the Euler constant. In the limit $q \to 0$, our
full expression (\ref{perturb_correl}) gives back the result of
Ref.~\cite{calabrese_fdr_randommass}. In the large time separation
limit $u \gg 1$, keeping $v$ fixed, one obtains the result:
\begin{eqnarray}
&&F_C(v,u) = \frac{1}{u} F_{C,\infty}(v) + {\cal O}(u^{-2})
  \label{1ou_decay} \\
&&F_{C\infty}(v) = A_{C \infty} v F_R^{\text{eq}}(v) \quad, \quad A_{C
  \infty} = 2 
  + 2\sqrt{\frac{6\epsilon}{53}} \nonumber \\\label{property_asympt}
\end{eqnarray}
This ${\cal O}(u^{-1})$ decay in (\ref{1ou_decay}) is expected
from RG arguments, and has been explicitly checked for different pure 
critical systems \cite{calabrese_review_fdr}.
However, this relation (\ref{property_asympt}) is a priori non trivial
and can not be obtained from general arguments. This 
identity was also found for the pure $O(N)$ model at 
criticality to one loop order \cite{calabrese_on_oneloop} as well as  
in the glass 
phase of the Sine Gordon model with random phase shifts 
\cite{schehr_co_prl} and it would be  
interesting to investigate whether such a behavior
(\ref{property_asympt}) can be obtained from more general arguments.    

The full expression for ${\cal C}^q_{tt_w}$ (\ref{perturb_correl}) also
allows to compute the structure factor ${\cal C}^q_{tt}$. It is
obtained from (\ref{perturb_correl}) in the limit $v \to 0$, $u \to 1$
keeping $v/(u-1) = q^z t$ fixed, and we check that one
recovers the previous result obtained in
Ref. \cite{kissner_random_mass}. Thus, one explicitly checks, at order
${\cal O}(\sqrt{\epsilon})$, that the dynamical exponent $z$ associated
to dynamical {\it equilibrium} fluctuations is the same as the
one associated to {\it nonequilibrium} relaxation.

As noticed previously for the response function, the large $v$
behavior of $F_C(v,u)$ is a power law $F_C(v,u)\propto
v^{-1}$. Therefore, given the scaling form (\ref{janssenscalingcorr}),
the computation of the autocorrelation ${C}_{tt_w}$ has to be
done carefully. Just as for the response, 
we thus compute the correlation function ${\cal
C}^q_{tt_w}$ from (\ref{Eq_C}) for any fixed $q$ and then perform the
Fourier transformation. One obtains the scaling form
\begin{eqnarray}\label{scal_auto_correl}
{C}^{}_{tt_w} = K_d 
\frac{A^0_{C} + A^1_{C}
\ln{(t-t_w)}}{(t-t_w)^{(d-2+\eta)/z}}\left(\frac{t}{t_w}\right)^{\theta}
{\cal F}(t/t_w)
\end{eqnarray}
with 
\begin{eqnarray}
&&A^0_{C} = 1 - \frac{1}{2}\sqrt{\frac{6\epsilon}{53}} +  \rho_{
 R} \quad, \quad A^1_{C} = \frac{1}{2}\sqrt{\frac{6\epsilon}{53}}
 \nonumber \\ 
&&{\cal F}(u) = \frac{1}{1+u} + {\cal O}(\epsilon)
\end{eqnarray}
The same remarks, concerning the response, 
made before Eq. (\ref{alter_scal_auto_resp}) also hold here for the
autocorrelation.

We now turn to the FDR, first in Fourier space. Given the scaling
forms for the response ${\cal R}^q_{tt_w}$ (\ref{janssenscalingresp})
and for the 
correlation ${\cal 
C}^q_{tt_w}$ (\ref{janssenscalingcorr}) we have explicitly checked
here, the FDR 
${X}^q_{tt_w}$  
takes the simple scaling form in the regime $q/\Lambda_0 \ll 1$:
\begin{eqnarray}
\left({X}^q_{tt_w}\right)^{-1} = F_{X}(q^z(t-t_w),t/t_w)
\end{eqnarray}
We have obtained the complete expression for the scaling function
$F_{X}(v,u)$, which at variance with the pure $\text{O}(N)$ model at
criticality is a function of both $q^z t$ and $q^z t'$. 
In the large time separation limit $u \gg 1$,
keeping $v$ fixed, one obtains, as a consequence of
(\ref{property_asympt}): 
\begin{eqnarray}\label{Xq}
\lim_{u \to \infty} \left({X}^q_{tt_w}\right)^{-1} = 2 +
\sqrt{\frac{6\epsilon}{53}} + {\cal O}(\epsilon) 
\end{eqnarray}
{\it independently} of $v$, {\it i.e.} of (small) wave vector $q$, which
coincides of course with the asymptotic value for the $q=0$ mode
obtained in Ref. \cite{calabrese_fdr_randommass}. We can check easily,
using the result of Ref. \cite{calabrese_on_oneloop},
that this property, independence on $v$ on the asymptotic limit, holds
also for the pure model 
at one loop, and it was also found in the 
glass phase of the Sine Gordon model with random phase shifts
\cite{schehr_co_prl}. 

As we saw previously, the large $v$ power law behavior of the scaling
function $F_R^{\text{eq}}(v)$ prevents us to use the argument of
Ref. \cite{calabrese_on_oneloop} for the present case. Therefore
on computes directly the FDR for the local correlation and
associated response $X^{x=0}_{tt_w}$. It is also characterized by a
scaling function of $t/t_w$, which can be simply written as
\begin{eqnarray}\label{local_FDR}
&&({X}^{x=0}_{tt_w})^{-1} = {\cal F}_{X}(t/t_w) \\
&&{\cal F}_{X}(u) = 2 \frac{u^2+1}{(u+1)^2} +
  \sqrt{\frac{6\epsilon}{53}} \left(\frac{u-1}{u+1}\right)^2 + {\cal
    O}(\epsilon)  
\end{eqnarray}
where ${\cal F}_{X}(u)$ is a monotonic increasing function of $u$, it
interpolates between $1$, in the
quasi-equilibrium regime for $u\to 1$, and its asymptotic value for $u
\to \infty$ given by
\begin{equation}
\lim_{t/t_w \to \infty} ({X}^{x=0}_{tt_w})^{-1} = \lim_{t/t_w \to \infty}
({X}^{q=0}_{tt_w})^{-1} = 2 + \sqrt{\frac{6\epsilon}{53}} + {\cal
  O}(\epsilon)   
\end{equation}
which shows explicitly, at order ${\cal O}(\sqrt{\epsilon})$ that the
asymptotic FDR for both the total 
and the local magnetization are indeed in the same.  
 
Let us next present results from our Monte Carlo simulations of the
 relaxational dynamics of the randomly diluted Ising model
 (\ref{eq_Hamil}) in dimension $d=3$, which were done on 
 $L\times L \times L$ cubic lattices with periodic boundary
 conditions. We first focus on the following situation where 
the system is initially prepared in a random initial
 configuration with zero magnetization $m_0 = 0$. At each time
 step, the $L^3$ 
 sites are then  
sequentially updated : for each site $i$, the move $s_i \to - s_i$ is
 accepted or rejected according to Metropolis rule.       
%
If one gradually decreases $p$ the fraction of magnetic sites will be reached
 below which the  
system no longer exhibits a transition to ferromagnetic order at any
finite temperature. This happens at the percolation threshold, for
 which $T_c(p_c) = 0$ \cite{ballesteros_tc,heuer_tc}. For different
 values of $p>p_c$, we compute
the spin-spin auto-correlation function defined as
\begin{equation}\label{def_correl_num}
{C}_{tt_w} = {\frac {1}{L^3}}  \sum_{i} \overline{\langle
 s_i(t)~s_i(t_w) \rangle} 
\end{equation}
In the following we will also be interested in the connected
correlation function $\tilde{C}(t,t_w)$ defined as
\begin{eqnarray}\label{def_connec}
\tilde C(t,t_w) = {\frac {1}{L^3}}  \sum_{i} \overline{\langle
 s_i(t)~s_i(t_w) \rangle  - \langle s_i(t)\rangle \langle s_i(t_w) \rangle}   
\end{eqnarray}
In order to obtain better statistics, ${C}_{tt_w}$ (or $\tilde
C(t,t_w)$) is averaged
over a suitably chosen time window $\Delta_t$ around $t$, with
$\Delta_t \ll t$. All our data are obtained for a lattice linear size
$L=100$, as 
an average over 500 independent initial conditions and 
disorder configurations. We also produced data (not shown here) for
the spatial correlation function, for the same system size, 
to ensure that our results are not influenced by finite size effects. 

\begin{figure}
\includegraphics[width=\linewidth]
{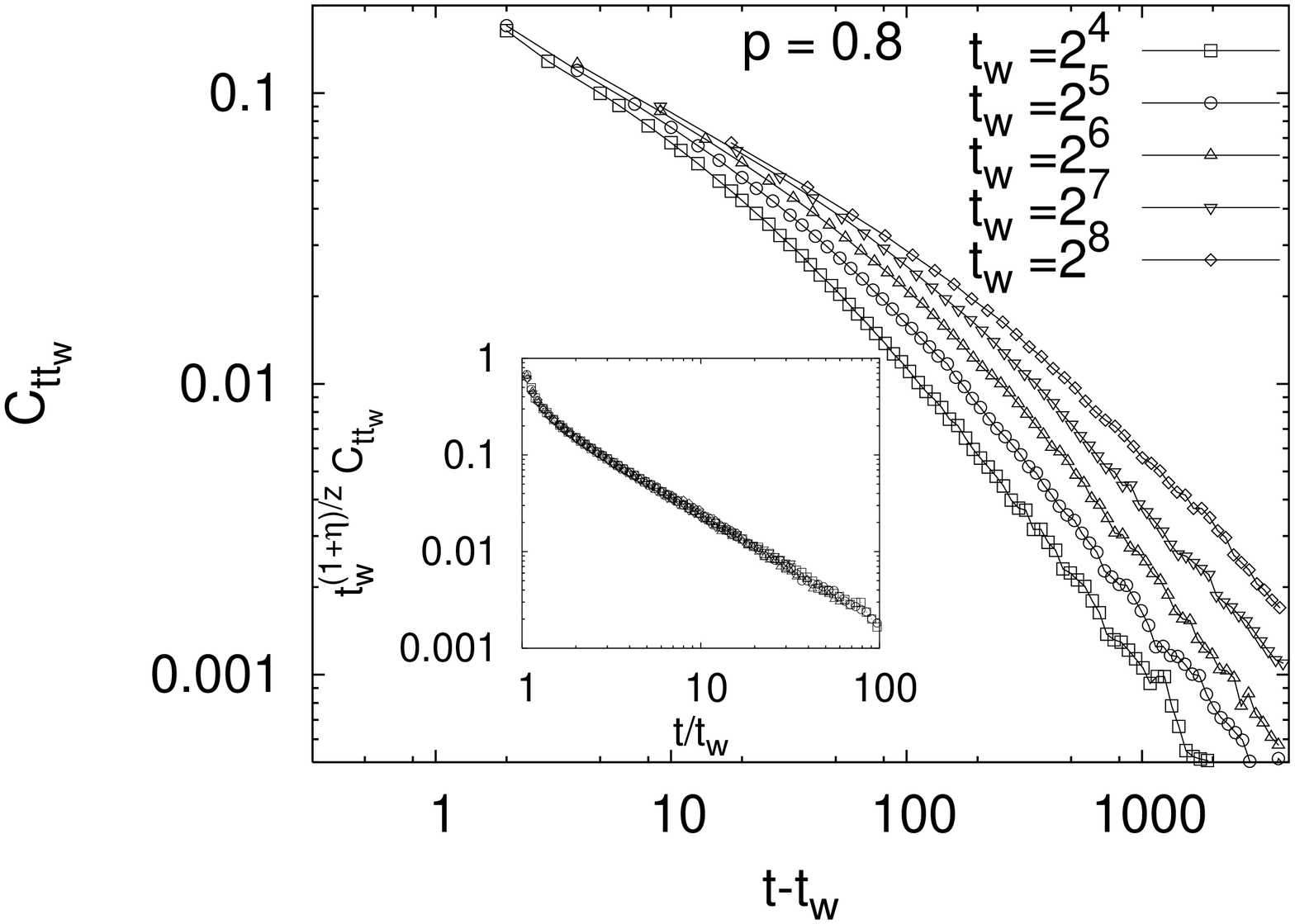}
\caption{Log-log plot of auto-correlation function ${\cal C}_{tt_w}$
 vs. $t-t_w$. {\bf Inset:} Scaling plot of ${\cal C}_{tt_w}$ as function
 of $t/t_w$. Here, the system is initially prepared in a random initial 
configuration with zero magnetization.} 
\label{fig1}
\end{figure}

Fig.~\ref{fig1} shows the auto-correlation function ${C}_{tt_w}$ as a
function of $t-t_w$ for different values of the waiting time 
$t_w = 2^4,~2^5,~2^6,~2^7$ and $2^8$ at $p=0.8$. One observes a clear
dependence on $t_w$, which indicates a non-equilibrium dynamical
regime. We have also checked that for this choice of initial
conditions, $C(t,t_w)$ and $\tilde C(t,t_w)$ do coincide. The scaling
form obtained from the RG analysis 
(\ref{scal_auto_correl}) suggests, discarding the logarithmic
correction, to plot $t_w^{(1+\eta)/z}{C}_{tt_w}$ as a function of
$t/t_w$. Taking the values $\eta = 0.0374$ from
Ref.\cite{ballesteros_tc} and $z=2.62$ from Ref.\cite{parisi_simu_rim},
we see in the inset of Fig.~\ref{fig1} that, for $p=0.8$, 
one obtains a good collapse of the curves for
different $t_w$. Notice that such scaling forms are also obtained in
more complicated disordered systems like $3$-dimensional spin-glasses
\cite{kisker_sg}. 

\begin{figure}
\includegraphics[width=\linewidth]
{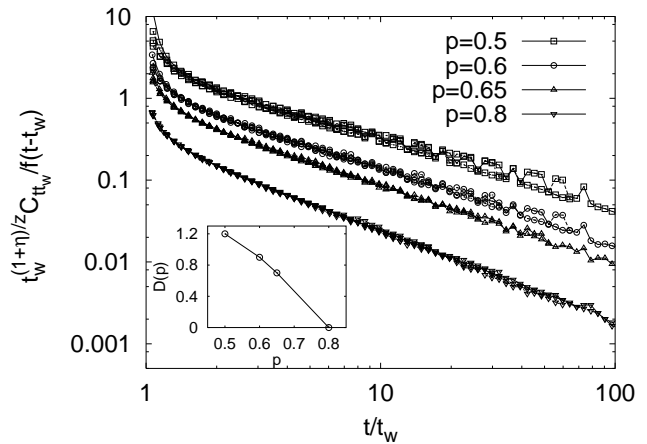}
\caption{$t_w^{(1+\eta)/z}{C}_{tt_w}/f(t-t_w)$ as a function of
 $t/t_w$ for different $p = 0.5,~0.6,~0.65$ and $0.8$. Waiting times
 $t_w$ corresponding to $p=0.5,~0.6,~0.65$ are
 $2^6,..,~2^{10}$ whereas for $p=0.8$,
 $t_w=~2^4,..,2^8$. $f(x)$ is defined in the text. 
{\bf Inset} : non-universal amplitude $D(p)$ as a function of $p$. 
Here, the system is initially prepared in a random initial 
configuration with zero magnetization.} 
\label{fig2}
\end{figure}

However, for different values of $p$, the best
collapse, under this form (\ref{scal_auto_correl}), would be obtained for a
$p$-dependent exponent  
$(1+\eta)/z$. Thus one would conclude that this exponent is
non universal \cite{heuer_tc}. 
Nevertheless, it is known \cite{parisi_simu_rim} that such
$p$-dependence occur due to corrections to scaling. Therefore, to include
them, we extend
the scaling form (\ref{scal_auto_correl}) as
\begin{equation}\label{correc_scal_auto_correl}
{C}_{tt_w} = \frac{1}{(t-t_w)^{(1+\eta)/z}}\left(
\tilde{F}_p\left( t/t_w \right) -  \frac{D(p)}{(t-t_w)^{b}}
\tilde{G}_p\left(t/t_w \right) \right)
\end{equation}
with $b=\omega_d/z$, where $\omega_d$ corresponds to the biggest
irrelevant eigenvalue of the RG in the dynamics, which is a priori
different from the leading corrections in the statics 
\cite{parisi_simu_rim}. Unfortunately, we do not
have any information on the function $\tilde{G}_p(x)$. We will thus
propose the simplest hypothesis, $\tilde{G}_p(x) = \tilde{F}_p(x)$. In
Fig.~\ref{fig2}, we show a plot of
$t_w^{(1+\eta)/z}{C}_{tt_w}/f(t-t_w)$, with $f(x) = 1 - D(p)x^{-b}$~:  
this results in a reasonably
good data collapse of the curves for different $t_w$, for $p =$0.5,
0.6, 0.65 and 0.8. For each value of $p$, this data collapse is obtained  via
the fitting of $3$ parameters : the exponents $b,z$ and the amplitude
$D(p)$. We found quite 
stable value of  
the exponents $z =2.6 \pm 0.1$ and
$b=0.23 \pm 0.02 $, which are {\it both independent} of 
$p$. Our value of $z$, together with 
$\omega_d = 0.61 \pm 0.06$ are consistent with
the value obtained 
by Parisi $et.~al$ \cite{parisi_simu_rim}. 
All the $p$-dependence is thus contained in the non universal
amplitude $D(p)$, as shown in the inset of Fig.~\ref{fig2}. According
to our data, the corrections to scaling in the quasi-equilibrium regime
vanish for $p=0.8$, {\it i.e.} $D(p=0.8) \simeq 0$, in agreement with a
previous numerical computation of the equilibrium
autocorrelation function \cite{heuer_tc}. Notice that this value
$p=0.8$ is also known \cite{ballesteros_tc}, in the statics, to 
minimize the corrections to scaling.   

\begin{figure}
\includegraphics[width=\linewidth]
{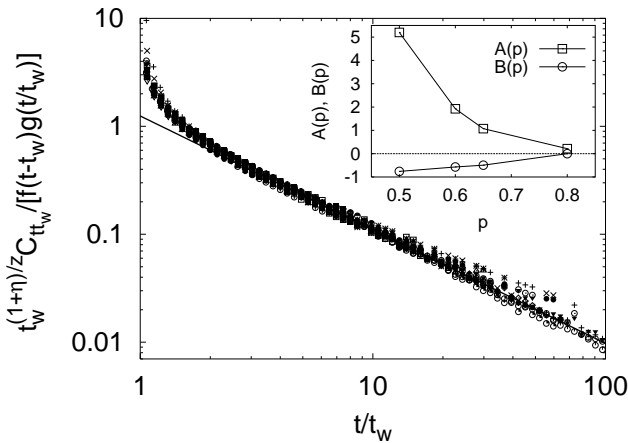}
\caption{Universality of ${C}_{tt_w}$ for $p=0.6, ~0.65, ~0.8$. The
 function $g(x)$ is defined in the text. {\bf
 Inset:} Non-universal amplitudes $A(p), ~B(p)$ as functions
 of concentration $p$. Here, the system is initially prepared in a random 
initial configuration with zero magnetization.}
\label{fig3}
\end{figure}

As shown on the log-log plot in Fig.~\ref{fig2}, and consistently
with the RG prediction (\ref{scal_auto_correl}),
$\tilde{F}_p(t/t_w)$ (\ref{correc_scal_auto_correl})
decays as a power law for $t \gg t_w$. However, this plot in
Fig.~\ref{fig2} would suggest that the decaying exponent depends,
namely decreases, with $p$. We expect instead that this
$p$-dependence is again due to corrections to scaling
\cite{parisi_simu_rim}. Consistently with the corrections 
we introduced in the quasi-equilibrium part of ${C}_{tt_w}$ in
Eq.~(\ref{correc_scal_auto_correl}), we propose the form
\begin{eqnarray}\label{correc_scal_F}
\tilde{F}_p (x) = A(p)x^{(1+\eta -\lambda_c)/z} \left(1 + B(p) x^{-b} \right) 
\end{eqnarray}   
where we (reasonably) assume that the dynamical corrections to scaling are
characterized by the {\it same}, $p$-independent, exponent $b = 0.23 \pm
0.02$
as obtained previously (\ref{correc_scal_auto_correl}). 
Therefore, for each value of $p$ one has three parameters to fit : the
exponent $\lambda_c/z$ and the amplitudes $A(p), B(p)$. We obtain a quite
stable fit for the different values of $p$, 
with the $p$-{\it independent} value of the decaying exponent
$\lambda_c/z$:
\begin{eqnarray}\label{exp_lambda}
\frac{\lambda_c}{z} = 1.05 \pm 0.03 
\end{eqnarray} 
all the $p$-dependence being contained in the non-universal amplitudes
$A(p), B(p)$ (see the inset in Fig.\ref{fig3}). As shown in
Fig.\ref{fig3}, the curves for different 
values of $p$ (and different $t_w$) in
Fig.~\ref{fig2} collapse on a master curve when we plot
$t_w^{(1+\eta)/z}{C}_{tt_w}/(f(t-t_w)g(t/t_w))$, with $g(x) =
A(p)(1+B(p)x^{-b})$, as a function of $t/t_w$. This fact supports
universality of the long-time 
non-equilibrium relaxation in this model. Our value for the exponent
$\lambda_c/z$, together with $z = 2.6 \pm 0.1$ gives for the initial
slip exponent $\theta' = 0.1 \pm 0.035$,
which is in rather good agreement with the two-loops RG result
$\theta'_{2\text{loops}} = 0.0868$
\cite{oerding_randommass_theta_twoloops}. Alternatively, this exponent 
could be measured by studying the initial stage of the relaxational dynamics
starting from a non-zero magnetization : this is left for future 
investigations \cite{raja_prep}.

Here also, one obtains that 
the corrections to scaling in Eq. (\ref{correc_scal_F}) vanish for
$p=0.8$. We notice that this result is in apparent contradiction with
the previous analysis of the 
non-equilibrium relaxation in this model performed in
Ref.\cite{parisi_simu_rim}, where 
the focus was on the non-connected susceptibility, a one-time quantity,
which instead claimed a ``perfect Hamiltonian'' for $p\simeq
0.63$. However, the statistical precision of our data does not allow us
to make a strong statement about this point, which
certainly deserves further investigations.

\begin{figure}
\includegraphics[width=\linewidth]
{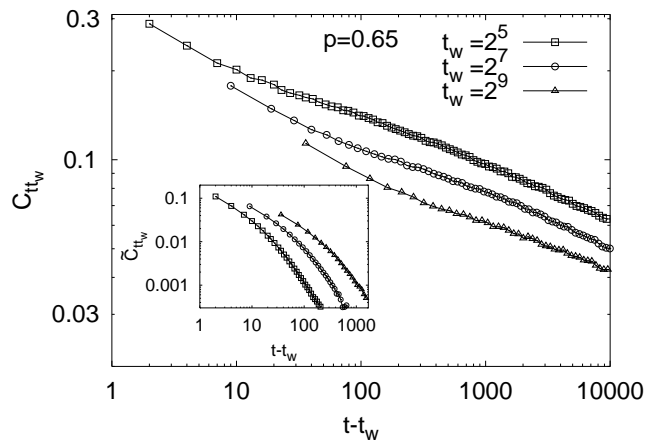}
\caption{ Log-log plot of the correlation $C(t,t_w)$ as a function of
$t-t_w$ for $p=0.65$. {\bf
 Inset:} Log-log plot of the connected correlation $\tilde C(t-t_w)$ 
as a function of $t-t_w$. In the inset, we use the same symbols 
as in the main figure. Here $m_0 = 1$.}
\label{fig4}
\end{figure}

\begin{figure}
\includegraphics[width=\linewidth]
{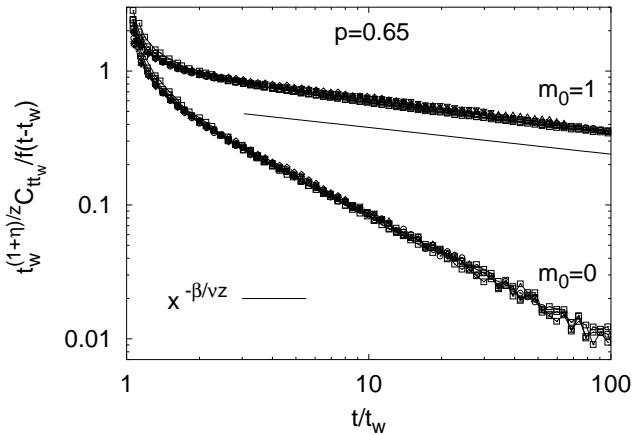}
\caption{$t_w^{(1+\eta)/z}{C}_{tt_w}/f(t-t_w)$ as a function of
 $t/t_w$ for $p=0.65$ and the two different initial conditions considered 
here, $m_0 =0$ and $m_0 = 1$. $f(x)$ defined in the text. The straight line
is a guide line for the eyes.}
\label{fig5}
\end{figure}

So far, we have focused on the relaxational dynamics occurring after a
quench from a completely disordered initial condition, with zero
initial magnetization $m_0 = 0$, to $T_c(p)$. 
But it is also interesting to study how these aging properties depend on
the initial conditions \cite{ci,berthier_xy,zheng_review}. We have
therefore performed 
numerical simulations where the system is initially prepared in a
completely ordered state:
\begin{eqnarray}\label{ord_ci}
S_i(t=0) = + 1 \quad, \quad \forall \quad \text{occupied site} \quad  i 
\end{eqnarray}
such that the initial magnetization is $m_0 = 1$. The system is then
quenched at $t=0$ to $T_c(p)$ 
and evolves according to the same aforementioned dynamical rules. 
We also compute the autocorrelation function $C(t,t_w)$ as defined in
Eq. (\ref{def_correl_num}). The result of this computation for
$p=0.65$ is shown on
Fig.~4, where we plot $C(t,t_w)$ as a function of $t-t_w$, for
different $t_w = 2^5,2^7,2^9$. Here also, one observes a clear dependence
on the waiting time $t_w$, which indicates that the system is aging. 
Notice however that, at variance with the previous situation (Fig. 1), the
correlation for a given $t-t_w$, decreases as $t_w$ increases. In
addition, at variance with the previous case $m_0 = 0$, the behaviors 
of the connected $\tilde C(t,t_w)$ (\ref{def_connec})
and the non-connected $C(t,t_w)$
correlations are qualitatively different : this is 
shown in the inset of Fig. \ref{fig4}, when one observes that $\tilde
C(t,t_w)$ decays indeed much faster \cite{foot_connected}. This
property could be relevant for the computation of the FDR in this
situation.   
The quantitative analysis of the correlation function $C(t,t_w)$ is
shown on Fig. \ref{fig5}. Indeed, the curves for different $t_w$ can
be plotted on 
a master curve if one plots, for different $t_w$,
$t_w^{(1+\eta)/z}{C}_{tt_w}/f(t-t_w)$ as a function of $t/t_w$. 
which suggests that also in that case the correlation function can be written 
under the scaling form as in Eq. (\ref{correc_scal_auto_correl}) with
$\tilde F_p(x) = \tilde G_p(x)$. However, 
as illustrated on Fig.~\ref{fig5}, the behavior of $C(t,t_w)$ is strongly
affected by the initial condition, the decay being much faster when
the system is initially in a random configuration with $m_0 =0$. 
More precisely, as
suggested on Fig.~\ref{fig5}, our
data for $m_0 = 1$ are compatible with the following scaling form
\begin{eqnarray}
&&{C}_{tt_w} \sim \frac{1}{(t-t_w)^{(1+\eta)/z}}\left(
1 -  \frac{D(p)}{(t-t_w)^{b}}\right)
\left(\frac{t}{t_w} \right)^{\frac{1+\eta}{2 z}} \nonumber \\
&& \sim t^{-\frac{\beta}{\nu z}} \quad, \quad t \gg t_w \label{scal_ord}
\end{eqnarray} 
where $\beta, \nu$ are the standard equilibrium critical exponents and
where we have used the hyperscaling relation $\beta/\nu =
(d-2+\eta)/2$. Thus, although we can
not show it analytically for the present problem, 
we believe that in that
case of a fully ordered initial condition ($m_0=1$), although the
system displays aging, the exponent
$\lambda_c$ is completely determined by the equilibrium exponents:
\begin{eqnarray}\label{lamb_ord}
\lambda_c = \frac{\beta}{\nu}
\end{eqnarray}     
This relation (\ref{lamb_ord}) can be understood by considering $C(t,0)$.
Indeed, for this particular initial condition (\ref{ord_ci}), one has
$C(t,0) = M(t)$, where $M(t)$ is the global magnetization at time $t$.
Therefore, at large time, from standard scaling argument 
$C(t,0) \sim t^{-\frac{\beta}{\nu z}}$, which thus gives the relation 
(\ref{lamb_ord}). Notice that this relation (\ref{lamb_ord}) is also 
found in the
context of pure critical point \cite{berthier_xy,zheng_review}.

To sum up, we have performed a rather detailed analysis of the
relaxational dynamics up to one loop of the randomly diluted Ising model
in dimension $d=4-\epsilon$. The computation of the correlation function
${\cal C}^q_{tt_w}$, including its associated scaling function, allows
us to show that the Fluctuation Dissipation Ratio reaches, in the large
time separation limit, a non trivial value $X^\infty$, independently of
small wave vector $q$. Although, due to the broad
relaxation time spectrum induced by the disorder, the standard argument
of Ref.~\cite{calabrese_on_oneloop} can not be applied here, we have 
performed an explicit computation in real space
which
shows explicitly that the limiting FDR associated to the total
magnetization, on the one hand, and the local one, on the other hand, do 
coincide. And in this respect, it would be interesting to further
investigate the FDR associated to other observables, like the energy for
instance \cite{calabrese_eff_temp,mayer_fdr}. These properties could also be
tested in numerical simulations. 

In addition, we have computed numerically, in $d=3$ the autocorrelation
function. It is characterized by a scaling form
fully compatible with our one loop RG calculation in real space. We have
however shown that this two times quantity is strongly 
affected by corrections to scaling, which remain to be understood 
more deeply from an analytical point of view. 
By taking them properly into account, our data 
suggest a universal, {\it i.e.} $p$-independent autocorrelation exponent
$\lambda_c$, which provides an ``indirect'' measurement of the initial
slip exponent $\theta'$, which is in reasonably good agreement with
two-loops RG prediction. Finally, we have shown that the critical dynamics 
following a quench from a completely ordered state ($m_0 = 1$) displays
also aging, but with a quantitative different behavior, the decaying 
exponent $\lambda_c$ being in that case completely determined by the 
{\it equilibrium} exponents.

\acknowledgments

GS acknowledges the financial support provided
through the European Community's Human Potential Program 
under contract HPRN-CT-2002-00307, DYGLAGEMEM and RP's work was
supported by the DFG (SFB277).


\begin{thebibliography}{35}


\bibitem{bray_domaingrowth_review}
A.J Bray,
\newblock { Adv. Phys.}, {\bf 43}, 357 (1994).

\bibitem{cugliandolo_fdr_pure}
L.~F. Cugliandolo, J.~Kurchan and G.~Parisi,
\newblock { J. de Phys. I}, {\bf 4}, 1641 (1994).

\bibitem{godreche_crit_ferro_review}
C.~Godr{\`e}che and J.M. Luck,
\newblock { J. Phys.: Condens. Matter}, {\bf 14}, 1589 (2002).



\bibitem{calabrese_review_fdr}
For a~recent review~see P.~Calabrese and A.Gambassi, cond-mat/0410357
	(2004). 


\bibitem{cugliandolo_leshouches}
L.~F.~Cugliandolo, {\it Dynamics of glassy
  systems} to appear in {\it Slow relaxation and nonequilibrium
  dynamics in condensed matter}, J.~L.~Barrat {\it et al.},
Springer-Verlag, 2002. 






\bibitem{godreche_fdr_crit_ferro}
C.~Godr{\`e}che and J.M. Luck,
\newblock { J. Phys. A}, {\bf 33}, 9141 (2000).

\bibitem{calabrese_on_oneloop}
P.~Calabrese and A.~Gambassi.
\newblock { Phys. Rev. E}, {\bf 65}, 066120 (2002); Phys.Rev. E {\bf 66},
	066101 (2002). 


\bibitem{calabrese_on_conserv}
P.~Calabrese and A.~Gambassi, Phys.Rev. E {\bf 67}, 036111 (2003).



\bibitem{leticia_teff}
L.~F.~Cugliandolo {\it et al.}, Phys. Rev. E, {\bf 55}, 3898 (1997). 

\bibitem{calabrese_eff_temp}
P.~Calabrese and A.~Gambassi, J.~Stat.~Mech: Theor. Exp. P07013
	(2004). 


\bibitem{mayer_fdr}
P.~Mayer {\it et al.}, Phys. Rev. E {\bf 68}, 016116 (2003).

\bibitem{heuer_tc}
	H.-O. Heuer, J. Phys. A: Math. Gen. {\bf 26} L333 (1993);
	J. Phys. A: Math. Gen. {\bf 26} L341 (1993). 



\bibitem{belanger_exp}
D.~P. Belanger {\it et al.}, J. de Physique Colloque C8, {\bf 49}, 1229
(1988). 


\bibitem{harris_criterion}
A.~B.~Harris,
\newblock { J. Phys. C}, {\bf 7}, 1671 (1974).

\bibitem{folk_rim_review}
For a~review see R. Folk {\it et al.},
\newblock { Physics Uspekhi}, {\bf 46}, 169 (2003).

\bibitem{ballesteros_tc}
	H.~G. Ballesteros, L.~A.~Fern\'andez, V.~Mart\'in-Mayor and
	A.~Mu\~noz~Sudupe, Phys. Rev. B {\bf 58}, 2740 (1998) and
	Ref. therein. 



\bibitem{foot_borel}
The summability of the perturbative expansions in diluted systems is a
	complex issue. For the present problem, it is discussed in
	\cite{folk_rim_review}. 


\bibitem{grinstein_randommass_z_oneloop}
G.~Grinstein {\it et al.},
\newblock { Phys. Rev. B}, {\bf 15}, 258 (1977).

\bibitem{janssen_randommass_z_twoloops}
H.~K.~Janssen {\it et al.},
\newblock { J. Phys. A}, {\bf 28}, 6073 (1995).

\bibitem{prudni_three_loops}
V.V. Prudnikov et~{al}, \newblock{JETP}, {\bf 87} 527 (1998).






\bibitem{parisi_simu_rim}
G.~Parisi {\it et al.},
\newblock { Phys. Rev. E}, {\bf 60}, 5198 (1999).

\bibitem{kissner_random_mass}
J.~G.~Kissner,
\newblock { Phys. Rev. B}, {\bf 46}, 2676 (1992).


\bibitem{oerding_randommass_theta_twoloops}
K.~Oerding and H.~K.~Janssen,
\newblock { J. Phys. A}, {\bf 28}, 4271 (1995).


\bibitem{andrei_ext}
A.~A.~Fedorenko, Phys. Rev. B, {\bf 69}, 134301 (2004).


\bibitem{calabrese_fdr_randommass}
P.~Calabrese and A.~Gambassi,
\newblock { Phys. Rev. B}, {\bf 66}, 212407 (2002).



\bibitem{janssen_gen}
H.K. Jansen {\it et al.}, 
\newblock{Z. Phys. B}, {\bf 73}, 539 (1989).


\bibitem{schehr_co_pre}
G.~Schehr and P.~{Le Doussal},
\newblock { Phys. Rev. E}, {\bf 68}, 046101 (2003).

\bibitem{chauve_erg}
P.~{Le Doussal} P.Chauve,
\newblock { Phys. Rev. E}, {\bf 64}, 051102 (2001).
\newblock cond-mat/0006057.

\bibitem{scheidl_multilocal}
S.~Scheidl and Y.~Dincer, cond-mat/0006048 (2000).

\bibitem{foot_univ}
up to an non-universal scale $q \to \lambda q$.

\bibitem{henkel_lsi}
M.~Henkel {\it et al.}, Phys. Rev. Lett. {\bf 87}, 265701 (2001).



\bibitem{schehr_co_prl}
G.~Schehr and P.~{Le Doussal}, Phys. Rev. Lett {\bf 93}, 217201 (2004).


\bibitem{kisker_sg}
J. Kisker {\it et al.}, Phys.Rev. B {\bf 53}, 6418 (1996).



\bibitem{raja_prep}
R.~Paul and G.~Schehr in progress.


\bibitem{ci}
A.~J.~Bray~{\it et al.}, Phys. Rev. B {\bf 43}, 3699 (1991);
A.~Picone~and~M.~Henkel, J. Phys. A {\bf 35}, 5575 (2002). 

\bibitem{berthier_xy}
L.~Berthier {\it et al.}, J.Phys. A {\bf 34}, 1805 (2001).

\bibitem{zheng_review}
For a review see B.Zheng, Int. J. Mod. Phys. B {\bf 12}, 1419 (1998). 

\bibitem{foot_connected}
Note also that the large time behavior of $\tilde C(t,t_w)$ also
depends on $m_0$. 


\end{thebibliography}

\end{document}